\documentclass[pre,floatfix,superscriptaddress,twocolumn,showpacs]{revtex4-1}
\usepackage{graphicx}
\usepackage{amsmath,amssymb,amsfonts}
\usepackage{mathtools}
\usepackage{bm}
\usepackage{color}
\usepackage[colorinlistoftodos]{todonotes}

\usepackage[toc,page]{appendix}
\makeatletter
\renewcommand{\@fnsymbol}[1]{%
  \ifcase#1\or †\or ‡\or *\or **\or \ddagger\or \mathsection\or \mathparagraph\or \|\or **\else\@arabic{#1}\fi}
\makeatother

\begin{document}

\title{Active Filaments on Curved Surfaces: From Single Filaments to Dilute Suspensions}

\author{Giulia Janzen}
\thanks{G.J. and E.D.M. contributed equally to this work.}
\affiliation{Department of Theoretical Physics, Complutense University of Madrid, Madrid, 28040, Spain}

\author{Euan D. Mackay}
\thanks{G.J. and E.D.M. contributed equally to this work.}
\affiliation{School of Life Sciences, University of Dundee, Dundee, DD1 5EH, United Kingdom}

\author{Rastko Sknepnek}
\email{r.sknepnek@dundee.ac.uk}
\affiliation{School of Life Sciences, University of Dundee, Dundee, DD1 5EH, United Kingdom}
\affiliation{School of Science and Engineering, University of Dundee, Dundee, DD1 4HN, United Kingdom}

\author{D. A. Matoz-Fernandez}
\email{dmatoz@ucm.es}
\affiliation{Department of Theoretical Physics, Complutense University of Madrid, Madrid, 28040, Spain}

\date{\today}

\begin{abstract}
Curvature plays a central organizational role in active polymer dynamics. Using large‑scale Langevin‑dynamics simulations, we study active semiflexible filaments confined to smooth curved surfaces and map how curvature, bending rigidity, and activity interact. We find geodesic alignment, curvature lensing, and curvature‑induced trapping. In particular, regions of negative Gaussian curvature localize filaments and hinder global surface exploration. These results show how surface geometry can be used to control the organization and transport of active matter on curved substrates
\end{abstract}
\maketitle
\section{Introduction}

Curvature is integral to the architecture and function of living systems, paradigmatic examples of active matter, across scales. During morphogenesis, epithelial sheets fold into tubes and furrows, generating structures such as the vascular system, brain and spinal cord~\cite{simon2004vascular,fatima2014murine,Whisler2023,vijayraghavan2017mechanics}. Surface geometry contributes to long-range coordination in the spiral drift of cells across the hemispherical cornea~\cite{kostanjevec2024spiral,schamberger2023curvature}. At the subcellular level, organelles such as mitochondria remodel their membranes during apoptosis~\cite{Chuai2006,Buske2011,Callens2020substrate,cereghetti2006many}. Understanding the role of curvature on active dynamics is, therefore, an important challenge ~\cite{metselaar2019topology,matoz2020wrinkle,mietke2019self,khoromskaia2023active,hoffmann2022theory,vafa2022active,wang2023patterning,Sknepnek2015,Henkes2018,Ehrig2017,bruss2017,castro2018active,Schonhofer2022,janssen2017aging,Iyer_2023}.

To investigate these effects in controlled conditions, synthetic systems have been developed using microtubules and ATP-powered kinesin motors confined to curved oil–water interfaces. In both spherical \cite{keber2014} and toroidal \cite{ellis2018} geometries, these systems exhibit rich curvature–activity coupling, such as oscillatory defect dynamics \cite{keber2014} and defect unbinding transitions \cite{ellis2018}. Notably, such phenomena can be captured by continuum models \cite{alaimo2017curvature,Henkes2018,pearce2019geometrical} that do not explicitly resolve the filamentous structure of the constituents.

However, little is known about how the interplay of activity and curvature affects long, semi-flexible chains, when the extended nature of the agents cannot be neglected and leads to new motion patterns. Only recently, experiments have demonstrated that confining cytoskeletal filaments to the inner surface of spherical lipid vesicles can induce robust collective motion patterns, such as streams, polar vortices, equatorial bands, and globally arrested states~\cite{Hsu2022}. While this work established a link between curvature and pattern selection, it was limited to spherical surfaces (i.e., surfaces of constant curvature), leaving open how spatially heterogeneous or sign-changing curvature affects the dynamics, and how filament flexibility and activity interact with the geometry.

Since experimental control of the vesicle shape and curvature distribution is not simple, computational models can be used to map out key properties of active filaments confined to move on curved surfaces. On flat substrates, numerical simulations of semiflexible filaments revealed interesting effects such as coiling and clustering, as well as nontrivial unwinding transitions \cite{jiang2014motion,IseleHolder2015,duman2018collective,Bianco2108,winkler2020physics,janzen2024active,winkler2025conformational}. Confinement to curved surfaces introduces additional elastic penalties, since passive filaments are generically misaligned with local geodesics \cite{Manning1987, Guven2012,guven2014environmental}. All these combine to form intricate physics that remains poorly understood.

To address how spatially varying curvature influences the dynamics of active semiflexible filaments, we perform large-scale Langevin dynamics simulations on closed curved surfaces with fixed geometry. Our aim is to go beyond spherical confinement and investigate how sign-changing curvature interacts with filament flexibility and activity to produce new dynamical regimes. By systematically varying surface shape from spheres to Gaussian bumps and necked Cassini ovals and exploring densities from isolated filaments to dilute suspensions, we disentangle the effects of curvature, stiffness, and collisions. We find that curvature can significantly guide filament dynamics, with activity and bending rigidity competing to determine alignment with geodesics. At the collective level, we observe curvature-induced confinement and the emergence of self-organized rotating bands. Notably, on Cassini surfaces, regions of negative curvature promote filament trapping and hinder global exploration. These results highlight surface geometry as a powerful mechanism to direct active filament organization.

The paper is organized as follows. Section~\ref{theory} introduces the theoretical framework for the bending energy of passive filaments on curved surfaces. Section~\ref{sec:simulation_model} describes the simulation model for active semiflexible chains. Section~\ref{sec:results} presents our main findings, focusing on both single-filament behavior and collective dynamics on curved surfaces. Finally, Section~\ref{sec:conclusions} summarizes our conclusions and outlines future directions.

\section{Overview of the behavior of filaments on curved surfaces}
\label{theory}
Point‐like active particles confined to a smooth, curved two‐dimensional surface embedded in Euclidean space exhibit nontrivial dynamics due to the effects of the surface curvature \cite{Sknepnek2015,Henkes2018,Schonhofer2022,Iyer_2023}. For example, the curvature induces a geometric torque on the particles, directing their motion along the geodesics \cite{paper-spin-connection}. 

In many experimental setups, however, active agents are better represented as semi-flexible or stiff filaments rather than point particles. The influence of activity on filament behavior remains only partially understood, even in the flat case~\cite{isele2015self, prathyusha2018dynamically, duman2018collective, winkler2020physics, Fazelzadeh2023,janzen2024density, janzen2024active}. For example, in contrast to passive filaments, where generic features of the system do not depend on the specifics of the discrete bond potentials, in the presence of activity, one needs to carefully distinguish between different types of discrete description of filaments \cite{winkler2025conformational}. 

For a filament embedded in a curved surface, in the absence of activity, its configuration is governed by intrinsic and extrinsic curvature; describing these effects presents a significant theoretical challenge  \cite{Manning1987,nickerson1988intrinsic,Guven2012,guven2014environmental,capovilla2002hamiltonians,Maria_Valencia_2019}. 
Theoretical approaches typically consider an inextensible filament modeled as a smooth curve \(\boldsymbol{\gamma}\) of length $L$ embedded in the smooth, orientable surface $\mathcal{S}$. The curve is parametrized by arclength \(s\), while the surface is parametrized by curvilinear coordinates $(u,v)$. Hence, $\boldsymbol{\gamma}(s)=\mathbf{r}(u(s),v(s))$, where $\mathbf{r}$ is the position vector in Euclidean space. At each point of the curve, one can assign the tangent vector $\mathbf{t}=\frac{\mathrm{d}\boldsymbol{\gamma}}{\mathrm{d}s}$ ($\left|\mathbf{t}\right| = 1$ since $s$ is the arclength parametrization), and define the surface unit normal $\mathbf{n}$. In general, $\mathbf{n}$ is not aligned with the normal to the curve, $\mathbf{N}=\frac{\mathrm{d}\mathbf{t}}{\mathrm{d}s}$. However, since $\mathbf{t}$ lies in the tangent plane of the surface, $\mathbf{n}\perp\mathbf{t}$ everywhere. One then defines the unit tangent normal $\mathbf{l}=\mathbf{t}\times\mathbf{n}$. The set of vectors $\{\mathbf{t},\mathbf{n},\mathbf{l}\}$ forms the Darboux frame \cite{docarmo76}, an embedded curve analogue of the Frenet-Serret frame for curves in $\mathbb{R}^3$. 

The scalar curvature of the curve is $k=\left|\frac{\mathrm{d}\mathbf{t}}{\mathrm{d}s}\right|$, i.e., it measures how the tangent vector changes its direction as one moves along the curve. The bending energy of the filament is given by~\cite{marko1995stretching}
\begin{equation}
    E = \frac{\alpha}{2} \int_{0}^{L} k^2(s) \, \mathrm{d}s,
    \label{eq:cont-energy}
\end{equation}
where \(\alpha\) is the bending rigidity. This expression is typically accompanied by the fixed-length requirement, e.g., by setting $|\mathbf{t}|=1$ everywhere, which is imposed using Lagrange multipliers. For an embedded curve, the curvature can be decomposed into the geodesic curvature, $k_g$, and the normal curvature, $k_n$, such that \(k^2 = k_g^2 + k_n^2\). The normal curvature measures the curvature of the surface in the direction of $\mathbf{t}$, while the geodesic curvature measures the intrinsic curvature of the curve, i.e., how much the curve bends along $\mathbf{l}$. 

Using the Darboux frame, the Euler–Lagrange equation that determines the equilibrium configurations of the filament is~\cite{nickerson1988intrinsic}, 
\begin{equation}
    k_g'' + k_g \left( \frac{k_n^2 + k_g^2}{2} - \tau_g^2 - c \right) - \frac{(k_n^2 \tau_g)'}{k_n} = 0,
    \label{eq:Euler-Lagrange}
\end{equation}
where \(\tau_g\) denotes the geodesic torsion, i.e., how the surface normal rotates around the tangent to the curve as one moves along the curve. Primes denote derivatives with respect to $s$, and the boundary conditions set the constant \(c\)~\cite{Manning1987,nickerson1988intrinsic,guven2014environmental,CastroVillarreal2019}. Assuming free ends \cite{nickerson1988intrinsic}, the boundary conditions at the filament endpoints (\(s = 0\) and \(s = L\)) are
\begin{equation}
    \begin{aligned}
    k_g(0) &= k_g(L) = 0, \\
    k_g'(0) &= 2k_n(0)\tau_g(0), \quad
    k_g'(L) = 2k_n(L)\tau_g(L).
\end{aligned}
\end{equation}

Equation \eqref{eq:Euler-Lagrange} is very hard to solve even on simple surfaces. It, however, highlights the coupling between the geometry of the surface and the elasticity of the filament, making even the passive case far from trivial.
A key result from~\cite{nickerson1988intrinsic} shows that geodesics (i.e., curves for which \(k_g = 0\)) are solutions to the Euler–Lagrange equation only when the condition \((k_n^2 \tau_g)' = 0\) is satisfied. Furthermore, a related theorem~\cite{Manning1987,nickerson1988intrinsic} states that on the plane or the sphere, an open-ended filament adopts an energy-minimizing configuration if and only if it follows a geodesic. 
This result follows from the fact that on both the plane (where for all curves \(k_n = 0\)) and the sphere (where \(k_n\) is constant), the geodesic torsion \(\tau_g\) vanishes identically, so that \((k_n^2 \tau_g)'=0\). However, under confinement, a closed loop on a sphere will deform away from the circular (geodesic) path, adopting a noncircular shape with increased bending energy~\cite{Guven2012}.

In general, the condition \((k_n^2\,\tau_g)'\! = 0\) is not satisfied, and the energy-minimizing configuration of the filament typically deviates from a geodesic~\cite{Manning1987,nickerson1988intrinsic,capovilla2002hamiltonians,guven2014environmental,CastroVillarreal2019}. Since our goal is to study active filaments on surfaces with nonuniform curvature, where bending energy, curvature, and self‐propulsion compete, the analytical treatment is not feasible and we instead resort to numerical simulations. We briefly note that recent work has shown that the mapping between the continuous and discrete behavior of active polymers is not necessarily straightforward \cite{winkler2025conformational}.

\section{Simulation Model} 
\label{sec:simulation_model}
We consider a collection of active filaments confined to a closed surface, with each filament composed of $N_\mathrm{b}$ beads. We ignore long-range hydrodynamic interactions, i.e., assume the dry limit \cite{MarchettiRev}.
The motion of bead $i$ is governed by the Langevin equation:
\begin{equation}
 m_i \ddot{\mathbf{r}}_i = -\gamma \dot{\mathbf{r}}_i + \mathbf{f}^{\mathrm{act}}_i + \sum_{j \neq i} \mathbf{f}_{ij} + \mathbf{R}_i(t),   
 \label{eq:motion}
\end{equation}
where $m_i$ is the bead mass, $\mathbf{r}_i$ is the position in Euclidean space of bead $i$ at time $t$, and $\gamma$ is the friction coefficient with the embedding surface. The active propulsion force $\mathbf{f}^{\mathrm{act}}_i=f_{\mathrm{p}}\mathbf{t}_i$ propels bead $i$ with magnitude $f_{\mathrm{p}}$ in the direction $\mathbf{t}_i$ set by the normalized sum of the unit-length vectors $\hat{\mathbf{r}}_{i+1,i}=(\mathbf{r}_{i+1}-\mathbf{r}_i)/\left|\mathbf{r}_{i+1}-\mathbf{r}_i\right|$ and $\hat{\mathbf{r}}_{i,i-1}=(\mathbf{r}_{i}-\mathbf{r}_{i-1})/\left|\mathbf{r}_{i}-\mathbf{r}_{i-1}\right|$ ~\cite{IseleHolder2015,Prathyusha2018}. The interaction between beads $i$ and $j$ is given as $\mathbf{f}_{ij} = -\nabla_{\mathbf{r}_i} \phi(r_{ij})$, where $r_{ij} = |\mathbf{r}_i - \mathbf{r}_j|$ and $\phi$ is the pair potential that describes both bonded and non-bonded interactions. $\mathbf{R}_i(t)$ represents thermal fluctuations and follows a Gaussian distribution with zero mean and variance $\langle \mathbf{R}_i(t) \cdot \mathbf{R}_j(t^\prime) \rangle = 4 \gamma k_\mathrm{B} T \delta_{ij} \delta(t - t^\prime)$ where $k_\mathrm{B}$ is the Boltzmann constant, $T$ is the temperature, and $\langle\dots\rangle$ denotes the thermal average.

Bonded interactions, $\phi_{\mathrm{B}}$, account for chain stretching and bending within the filament, with stretching described by a tether bond potential \cite{Noguchi2005} and bending by a harmonic angle potential \cite{Prathyusha2018}. Nonbonded interactions, $\phi_{\mathrm{NB}}$, capture steric repulsion and are modeled using the Weeks-Chandler-Anderson (WCA) potential \cite{weeks1971role}, \(\phi_{\textrm{WCA}}(r_{ij})=4\varepsilon\left[\left(\frac{\sigma}{r_{ij}}\right)^{12}-\left(\frac{\sigma}{r_{ij}}\right)^{6}+\tfrac{1}{4}\right]\) for $r_{ij}<2^{1/6}\sigma$, where \(\varepsilon\) sets the interaction energy scale and $\sigma$ is the monomer diameter. %

Surface confinement is enforced by projecting bead positions onto the surface and restricting velocities and forces to the local tangent plane. For a general smooth surface $\mathcal{S}$ parametrized as $\mathbf{r}(u,v)$, a point $P\in\mathbb{R}^3$ can be projected on $\mathcal{S}$ by minimizing $\psi(u,v)=\frac{1}{2}\left|\mathbf{r}(u,v)-P\right|^2$, e.g., by using Newton minimization with gradient $\nabla g=\left[(\mathbf{r}(u,v)-P)\frac{\partial\psi}{\partial u},(\mathbf{r}(u,v)-P)\frac{\partial\psi}{\partial v}\right]^T$. Projecting vector $\mathbf{a}$ onto the tangent plane is simply, $\hat{P}^T_{\mathbf{n}_i}(\mathbf{a})=\mathbf{a} - \left(\mathbf{a}\cdot\mathbf{n}_i\right)\mathbf{n}_i$, where $\mathbf{n}_i$ is the unit-length surface normal at $\mathbf{r}_i$. 

In this study, we simulate polymers with a degree of polymerization \(N_b = 80\), and work in the set of units where the monomer size \(\sigma = 1.0\), interaction energy scale \(\epsilon = 1.0\), thermal energy \(k_\mathrm{B} T/\varepsilon = 0.1\), and monomer mass \(m = 1.0\). Although our simulations are conducted using Langevin dynamics, we do not expect the qualitative behaviors discussed below to differ significantly under Brownian dynamics. On flat surfaces, inertial effects have primarily been shown to influence filament compactness~\cite{janzen2024density,Fazelzadeh2023}. Each polymer has a length \(L \approx b_{\textrm{bond}} \,(N_b - 1)\,\sigma,\) where \(b_{\textrm{bond}} = 0.86\). 
For simulations of filament suspensions, the systems were first equilibrated for $10^5$ time units using a time step $dt=10^{-4}$ before data collection. All simulations were performed using the GPU-accelerated SAMoS molecular dynamics package~\cite{SAMoS2024} with the BAOAB Langevin integrator~\cite{leimkuhler2015molecular}. Data were analyzed using custom Python scripts and visualized with ParaView~\cite{fabian2011paraview}.

We introduce a geometry-dependent dimensionless Péclet number defined as \( Pe = \frac{f_\mathrm{p} l_s} {\kappa} \), where \(l_s\) represents the characteristic length scale along the surface over which the effects of curvature become significant to a filament's motion. Specifically, we define \(l_s = 1/\sqrt{\left|K\right|_{\mathrm{max}}}\), where \(\left|K\right|_{\mathrm{max}}\) is the maximum of the absolute value of the Gaussian curvature. \(\kappa\) is the discrete equivalent of the continuum bending rigidity $\alpha$. This Péclet number quantifies the impact of surface curvature on filament dynamics. This approach is compatible with findings on spherical surfaces, where curvature introduces an additional timescale, $\tau = \frac{R}{v_0}$, with $R$ being the sphere's radius and $v_0$ the self-propulsion speed of the active particles. This timescale is in addition to the conventional rotational diffusion time and affects the dynamics of active Brownian particles on curved surfaces~\cite{Iyer_2023}.

\section{Results and discussion}
\label{sec:results}
\subsection{Single Active Filaments on a Gaussian Bump}
\label{sec:single_active}
\begin{figure}[t]
\includegraphics[width=1\columnwidth]{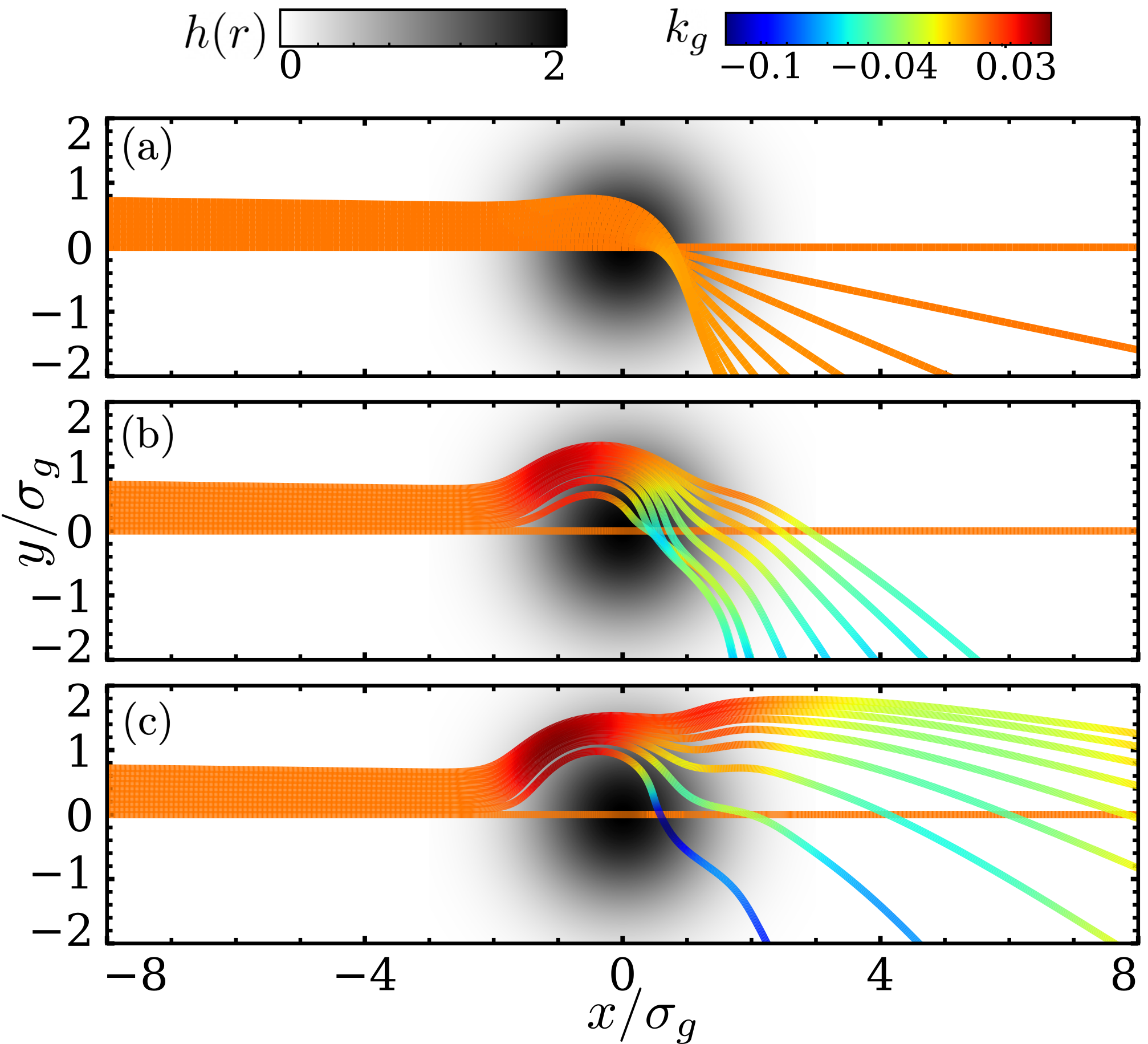}
\caption{\label{fig1} Deviation of active filaments from geodesics at activity \(f_\mathrm{p} = 10^{1}\) as a function of bending rigidity \(\kappa\) on a Gaussian bump with amplitude \(A=2\) and width \(\sigma_g=1\). Each panel shows trajectories of the head monomer of eight filaments initially placed along geodesics at different \(y\)-coordinates in the \(x\)–\(y\) plane. Trajectories are color-coded by the instantaneous geodesic curvature \(k_g\) averaged over filament length, and the grayscale background indicates the Gaussian bump height profile \(h(r)\). The geodesic curvature is computed numerically, and details of the numerical procedure are provided in Appendix~\ref{geodesic_normal}.
(a) For zero bending rigidity (\(\kappa=0\)), filaments closely follow their initial geodesics, with \(k_g\) ranging from \(10^{-8}\) to \(10^{-4}\).
(b) At \(\kappa=5\times10^{1}\) (\(Pe=10^{-1}\)), only the filament initialized at \(y=0\) remains on its geodesic, which has zero geodesic torsion (\(\tau_g=0\)) and minimizes bending energy. The others deviate from their initial geodesics and do not transition to alternative geodesics, as shown by the marked increase in \(k_g\) when crossing the bump.
(c) At high bending rigidity (\(\kappa=10^2\), \(Pe=5 \times 10^{-2}\)), all filaments except the one at \(y=0\) exhibit substantial deviation from geodesics. The higher \(k_g\) indicates that bending energy dominates filament dynamics, driving them away from geodesic paths.}
\end{figure}
We begin by studying the deterministic (i.e., noiseless) active dynamics of a single active filament confined to a Gaussian bump. The Gaussian bump has non-uniform curvature, with positive curvature at the apex and a narrow band of negative curvature at its base (Appendix \ref{Gaussican_curv}). The curvature also quickly decays to zero as one moves away radially, making it a good model to study the effects of non-uniform curvature in a flat background. In addition, a generic curve has both nonconstant normal curvature and nonzero geodesic torsion, so the filament configurations that minimize bending energy are typically not geodesics. 

The surface is axisymmetric about the vertical axis through its peak and can naturally be parameterized by \((r, \theta)\), where \(r\) measures the radial distance from the axis of symmetry and \(\theta\) is the azimuthal angle. The position vector \(\mathbf{r}(r,\theta) = \big(r\cos\theta,\, r\sin\theta,\, h(r)\big)\), describes points on the surface, with a height function \(h(r) = A\exp\left(-\frac{r^2}{2\sigma_g^2}\right)\), where \(A\) is the height at the peak, and \(\sigma_g\) sets the width of the bump. The height function \(h(r)\) describes the elevation above the \(xy\)-plane in Euclidean space \(\mathbb{R}^3\). This sets $l_s = \sigma_g^2/A$.

We initialize filaments sufficiently far from the bump that the Gaussian curvature can be neglected by placing them along geodesics (i.e., straight lines) along $x$ at different \(y\)-positions and track their trajectories under varying activity. If geodesics minimize the bending energy, the filaments remain on them, with geodesic curvature \(k_g\) remaining zero as the filament evolves. This can be understood as follows. Active force is tangent to the filament and the surface at every point, and as such, forces beads to follow geodesics\cite{paper-spin-connection}. Bending energy is minimized by remaining on a geodesic, and any deviation from it results in a restoring force that pushes each bead back to geodesics. Therefore, the entire filament follows a geodesic. 
For fully floppy filament, i.e., \(\kappa=0\), there is no bending penalty and the filament trajectory is fully determined by active force, i.e., it follows the initial geodesic it was placed on~\cite{paper-spin-connection} with \(k_g \approx 0\)  (Fig.~\ref{fig1}a). Note that due to the spatial discretization \(\Delta s = 1\) (Appendix \ref{discrete_vs_continuum}), the computed \(k_g\) fluctuates between \(10^{-8}\) and \(10^{-4}\), but decreasing \(\Delta s\) reduces these fluctuations.

As \(\kappa\) increases, filaments deviate from their initial geodesics and \(k_g\) increases (Fig.~\ref{fig1}b). 
Finally, for very stiff filaments, these deviations and the associated \(k_g\) become more pronounced (Fig.~\ref{fig1}c). Only the filament initialized at \(y=0\) continues to follow its geodesic because this geodesic has \(\tau_g=0\), corresponding to a bending energy minimum (Sec.~\ref{theory}). In this regime, bending energy dominates, and filaments only follow geodesics that also minimize this energy, as with the central geodesic. These observations highlight that tangential active forces drive filaments along geodesics. When activity is strong enough to overcome bending rigidity, filament dynamics are dictated by surface geometry, leading to alignment with geodesic paths.

We can understand this competition between activity and bending rigidity through a simple scaling argument. We move to the overdamped limit of the filament dynamics and compare the time taken for the filament to pass over the bump, driven by its active self-propulsion $t_a$, with the time taken for the filament to deform due to its elastic bending forces $t_b$. When $t_a\ll t_b$, we expect the filament to have passed smoothly over the bump on a geodesic trajectory before the elastic force has time to alter its dynamics. In Appendix \ref{sec:scaling}, we estimate the ratio $\frac{t_b}{t_a}\approx Pe \frac{A}{L}$. This means at large $Pe$ we expect $t_a\ll t_b$, so the filament is expected to remain on a geodesic, consistent with our numerical results.

\subsection{Motion on surfaces with spherical topology}
\label{sec:spher-topo}
\begin{figure}[t]
\includegraphics[width=1\columnwidth]{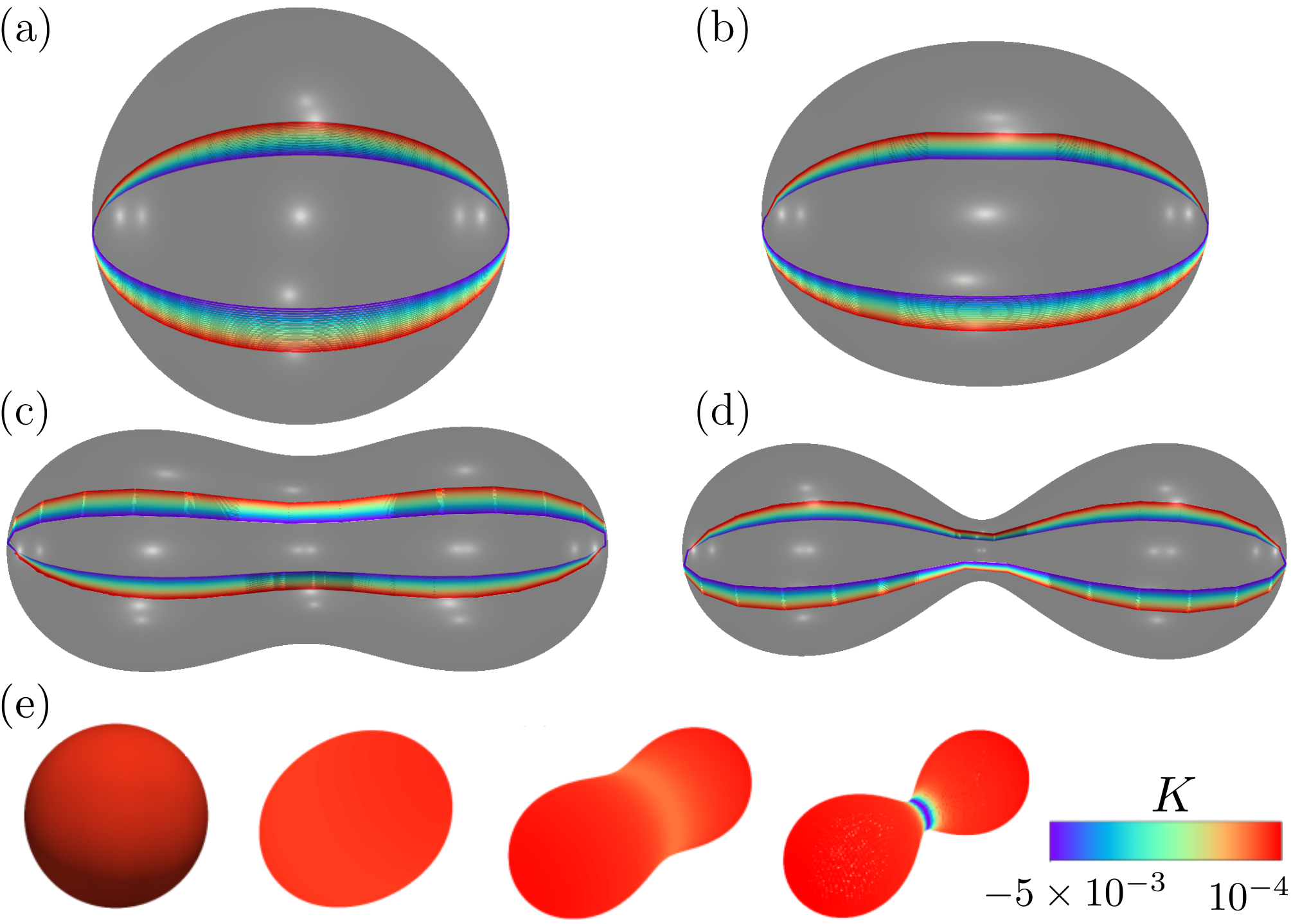}
\caption{\label{fig_cassini_non_int}Trajectories of the head monomers of eight non-interacting active filaments, each initially positioned along a different great circle on a sphere. Each filament is shown in a distinct color: the blue trajectory corresponds to the filament initialized on the equatorial great circle (i.e., the one passing through the center of the sphere), while the others lie on parallel great circles below the equator. As the surface gradually deforms, the filaments continue to follow geodesic paths.  
(a) Filaments on an undeformed spherical surface (\(a = 0\)).  
(b) Filaments on a prolate spheroid (\(a = 0.5b\)).  
(c) Filaments on a peanut-shaped surface (\(a = 0.9b\)).  
(d) Filaments on a peanut-shaped surface with a narrow channel (\(a = 0.99b\)).  
(e) Gaussian curvature $K$ on Cassini ovals for increasing values of \(a\). Surfaces are color-coded by Gaussian curvature, with the formula for \(K\) given in Appendix~\ref{Gaussican_curv}.}
\end{figure}
\begin{figure*}[t]
\includegraphics[width=18cm]{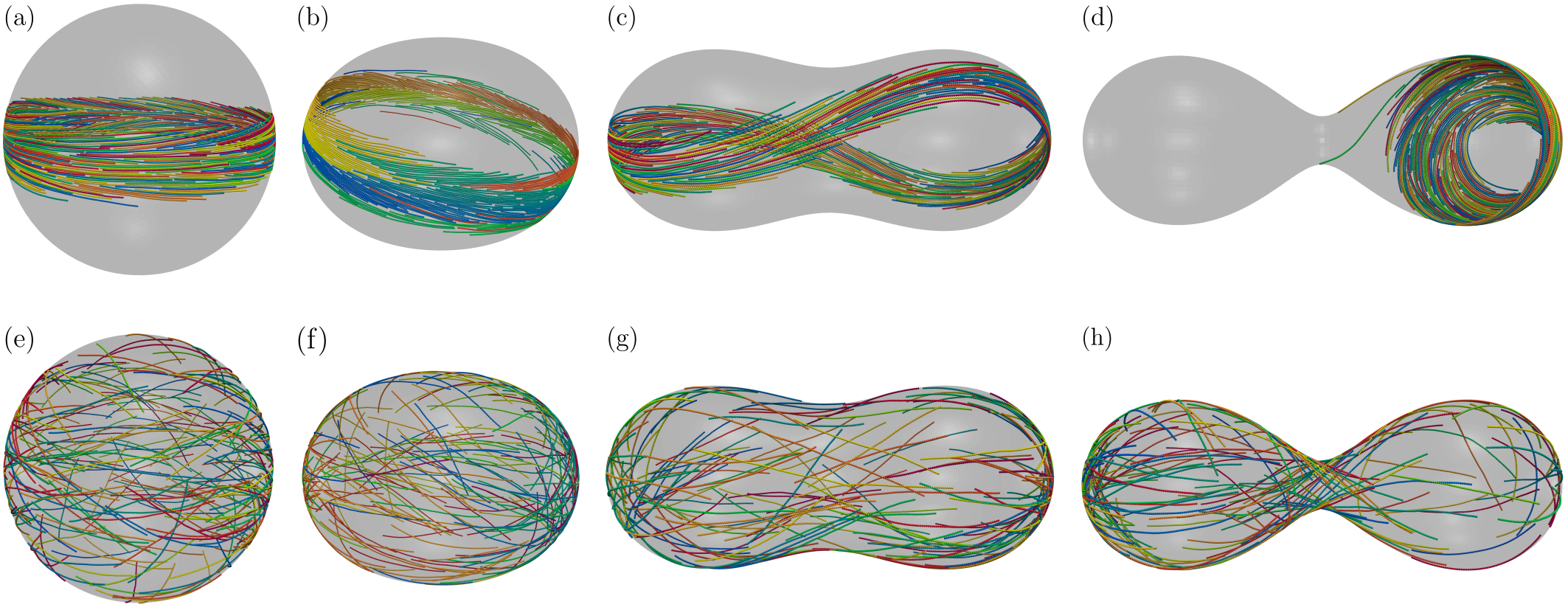}
\caption{\label{fig2} Steady-state configurations of stiff active filaments $\kappa=100$ at intermediate activity for four values of the Cassini oval parameter \(a\): (a) Sphere (\(a=0\)), (b) Egg-like shape (\(a=0.5b\)), (c) Peanut-like shape (\(a=0.9b\)), and (d) Peanut-like shape with a narrow channel (\(a=0.99b\)). Snapshots correspond to an intermediate density of \(0.16\), with filaments individually color-coded. On spherical-like surfaces (panels (a)–(c)), a rotating band state emerges. In contrast, for the narrow-channel peanut (panel (d)), where the curvature gradient between lobes is pronounced, filaments become trapped on one side.
(e)–(h): Single-filament dynamics initialized from the corresponding configurations in panels (a)–(d). Filament colors match those in the first row for visual consistency.}
\end{figure*}
Motivated by experiments~\cite{Hsu2022}, we focus on surfaces with spherical topology. 
A key distinction compared to the Gaussian bump is that the sphere is compact, and geodesics are closed loops, i.e., great circles. The geodesic torsion $\tau_g=0$, and an open active filament will follow a great circle (Fig.~\ref{fig_cassini_non_int}a). 
It, however, remains unclear how filament dynamics are affected if the spherical topology is retained but the surface is deformed to have non-uniform curvature. To address this question, we gradually deform the sphere to introduce spatial variations in curvature while preserving its closed topology. Specifically, we study the behavior of single active filaments on Cassini ovals revolved around the \(x\)-axis. These surfaces are implicitly defined by
\begin{equation*}
    f(x, y, z) = (x^2 + y^2 + z^2)^2 - 2a^2(x^2 - y^2 - z^2) + a^4 - b^4 = 0,
    \label{eq:cassini-ovals}
\end{equation*}
where the parameters \(a\) and \(b\) control the shape. In the limit \(a=0\), this equation reduces to a sphere of radius \(b\). For \(a \sim \frac{b}{2}\), the surface resembles a prolate spheroid, while for \(a > \frac{b}{2}\), it transitions to a peanut-shaped geometry with two lobes connected by a neck of negative curvature (Appendix \ref{Gaussican_curv}). For a sphere (\(a = 0\)), the natural length scale is simply the radius (i.e., \(l_s = b\)). For other Cassini ovals (\(a > 0\)), the maximum value of the Gaussian curvature, \(\left|K\right|_{\mathrm{max}}\), can be computed numerically. As shown in Fig.~\ref{fig_cassini_non_int}e, for peanut-shaped surfaces this maximum typically occurs in the neck region. The corresponding curvature values in this area are indicated in the color bar of Fig.~\ref{fig_cassini_non_int}e.

To explore the stability of geodesic trajectory, we gradually tune the surface shape by tuning values of $a$. We begin by placing a filament along a great circle on the sphere (\(a=0\)) as a reference case. After allowing it to complete a full revolution, we incrementally deform the surface by increasing \(a\) up to \(a=0.99b\), repeating the simulation and considering geodesic curvatures \(k_g \leq 10^{-12}\) effectively zero. Starting from a sphere, increasing \(a\) causes the filament to continue following a geodesic. At \(a=0.5b\) (prolate spheroid, Fig.~\ref{fig_cassini_non_int}b), \(k_g\) and \(\tau_g\) remain near zero, indicating it follows a geodesic that also minimizes bending energy. This remains true at \(a=0.9b\) and \(a=0.99b\) (peanut-shaped surfaces, Figs.~\ref{fig_cassini_non_int}c,d). In the most deformed case (Fig.~\ref{fig_cassini_non_int}d), \(k_g\) and \(\tau_g\) increase slightly but stay very small, suggesting the filament remains close to a geodesic. This slight increase is likely due to numerical errors.
  
Having established how curvature variations, activity, and bending rigidity govern the dynamics of individual filaments, we now turn to the collective behavior of many interacting active filaments confined to curved surfaces. This shift is motivated by experimental observations of coordinated motion on closed geometries, where interactions become central to the emergent dynamics~\cite{Hsu2022}. While our earlier analysis focused on trajectories initialized along specific geodesics, it is important to note that filament motion on a Cassini oval is also sensitive to initial conditions. We revisit this point at the end of the section, demonstrating how different initializations can lead to distinct patterns of spatial exploration. To connect with experimental relevance and gain insight into the collective behaviors, we begin by examining suspensions of active filaments.


Previous studies of dry active particles with polar or nematic alignment and short-range repulsion on spherical surfaces have shown that these systems can spontaneously form rotating polar bands around the equator~\cite{Sknepnek2015,Henkes2018,paper-spin-connection}. This collective behavior arises from a balance between the particles’ tendency to follow geodesic paths (i.e., great circles) and their mutual steric repulsion, with band width narrowing as activity increases.
Building on our analysis of single-filament dynamics on curved spherical surfaces, we now explore the collective behavior of stiff active filaments at intermediate densities. In this regime, the filaments self-organize into a rotating band near the equator (Fig.~\ref{fig2}a), closely mirroring the dynamics reported in prior simulations of active particles on curved geometries~\cite{Sknepnek2015,Henkes2018,paper-spin-connection}.
Notably, recent experiments~\cite{Hsu2022} have observed analogous phenomena: actin filaments gliding along the inner surface of spherical vesicles exhibit surface-density-dependent transitions. As the filament concentration increases, these systems develop polar bands and display two additional dynamical regimes: polar vortices at off-equatorial latitudes at intermediate coverage, and disordered filament clusters at high coverage. Our simulations, performed under comparable conditions, reproduce the emergence of equatorial polar bands at low densities, in agreement with these experimental observations. We do not not observe the additional dynamical regimes, as this modeling approach is not designed to explore the high coverage states where hydrodynamics are likely to be an important factor. 

To understand how non-uniform curvature affects collective dynamics, we further explore filament organization on Cassini ovals by varying the parameter \(a\). For \(a=0.5b\), the surface has the shape of a prolate sphenoid, and similar to their behavior on a sphere, filaments form a rotating polar band (Fig.~\ref{fig2}b). A comparable pattern emerges at \(a=0.9b\), where the surface becomes peanut-shaped (Fig.~\ref{fig2}c).
However, at \(a=0.99b\), the neck connecting the lobes narrows substantially, creating a pronounced curvature gradient. In this regime, filaments tend to become trapped on one side of the surface (Fig.~\ref{fig2}d). As a result, the equatorial polar band no longer forms, and motion becomes localized to a single lobe. Note that due to noise in the simulations, the filament configurations shown in Fig.~\ref{fig2}a--d do not exactly follow geodesics (Appendix \ref{Single_filament_on_Cassini}).

As discussed in Sec.~\ref{sec:single_active}, some geodesics on Cassini ovals with \(a>0\) resemble those on the sphere, so the similar filament organization observed for \(0<a \leq 0.9b\) is not surprising. In contrast, the qualitative change at \(a=0.99b\), where filaments become confined to one lobe, suggests a transition in the dynamics. This raises the question of whether the behavior in Fig.~\ref{fig2}d is driven solely by surface curvature or also influenced by crowding. 

To address this, we analyzed the dynamics of single filaments initialized with the same configurations as in Fig.~\ref{fig2}a-d. As shown in Fig.~\ref{fig2}e-h, single filaments explore the entire surface without preferring specific regions. We verified that this single-filament behavior persists even in the absence of noise (see Appendix~\ref{Single_filament_on_Cassini}). This indicates that, in the absence of interactions, individual filaments do not favor any particular area.  
These results suggest that the trapping observed at \(a=0.99b\) in Fig.~\ref{fig2}d does not arise from the intrinsic dynamics of single filaments, but likely results from crowding effects.

Overall, we conclude that in the dilute regime, filaments on the Cassini oval move collectively across a wide range of the parameter $a$. Notably, the flocking behavior observed on curved spherical surfaces~\cite{Hsu2022} is absent in the planar case at the same density~\cite{janzen2024density}, suggesting that curvature promotes flocking in this dilute regime, consistent with findings for polar and nematic active systems~\cite{Sknepnek2015,Henkes2018,paper-spin-connection}.
\begin{figure}[t]
\includegraphics[width=1\columnwidth]{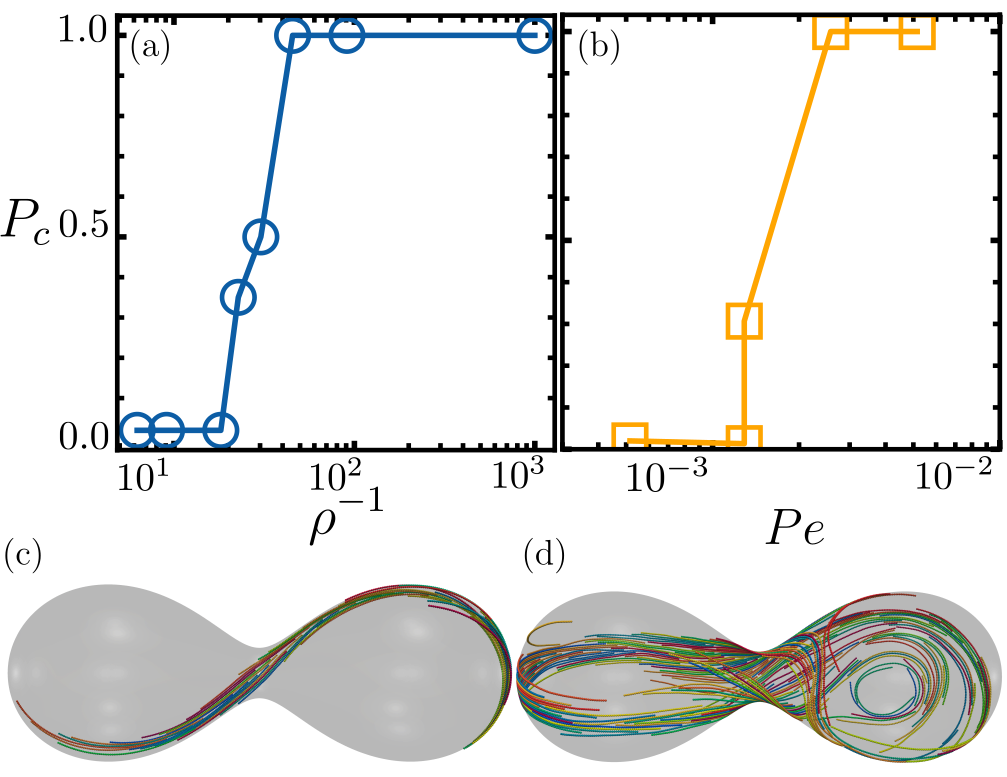}
\caption{\label{fig_trapping} Trapping–escaping transition. (a) Probability of crossing \(P_c\) as a function of inverse density \(\rho^{-1}\). At low densities (\(\rho^{-1} < 4.5 \times 10^{1}\)), filaments are trapped on one side of the peanut-shaped surface. At higher densities, all filaments escape and explore both lobes. (b) Probability of crossing \(P_c\) as a function of Péclet number \(Pe\) at fixed density \(\rho=0.16\). The behavior is similar to panel (a): at low \(Pe\), filaments are trapped, while increasing \(Pe\) enables exploration of both lobes. (c) Band-like behavior observed at \(\rho^{-1}=2.2 \times 10^{1}\); similar dynamics occur at higher densities. This configuration matches that seen for \(a=0.9b\) (see Fig.~\ref{fig2}c). For \(\rho^{-1}<4.5 \times 10^{1}\), behavior resembles Fig.~\ref{fig2}d. (d) Snapshot at \(\rho=0.16\) and \(Pe=8 \times 10^{-3}\). Unlike lower densities, filaments explore the full surface but do not form a coherent rotating band.}
\end{figure}

\subsection{Trapping effect as a function of density}
To further investigate the trapping effect observed for \(a = 0.99b\), we compute the probability that filaments cross the neck between the two lobes (Fig.~\ref{fig_trapping}). In the ``peanut-shape'' geometry, the left lobe corresponds to negative \(x\) values, the right lobe to positive \(x\), and the neck is located at \(x=0\). To quantify the crossing probability, we first let the system equilibrate and then monitor it over a long observation window of $t=10^5$ time units. This duration is chosen to ensure that filaments have sufficient time to explore the surface, so that those which appear confined are truly trapped. During this period, we count how many filaments change the sign of their  $x$-coordinate at least once. The crossing probability is defined as the fraction of filaments that exhibit such a sign change. Since we compare systems with different densities, for \(\rho<0.16\) we average over multiple initial conditions to match the total number of filaments analyzed in the highest-density case (\(\rho=0.16\)).

Measurements were performed at fixed Péclet number (\(Pe=10^{-3}\)) across densities from \(\rho=0.16\) down to \(\rho=10^{-3}\), i.e. single filament (Fig.\ \ref{fig_trapping}a). We observe that $P_c$, and thus surface exploration, increases as density decreases. In other words, lower densities promote broader exploration.   
At low densities, filaments form band-like trajectories (Fig.~\ref{fig_trapping}c) similar to those on the peanut-shaped surface (Fig.~\ref{fig2}c), collectively crossing the neck. These findings suggest that at this Péclet number, filaments tend to move cohesively. When density is high and the neck is too narrow to accommodate all filaments, they remain confined to a single lobe. At lower densities, collective neck crossing becomes more likely, allowing exploration of the full surface.
Moreover, we measured the time required for a filament to cross to the opposite lobe. When crossings occur, this time remains approximately constant across densities, with an average value of  \(t_c = 2.5 \times 10^3\) time units.

At fixed density (\(\rho=0.16\)), as $Pe$ increases, the probability of channel crossing $P_c$ increases, allowing filaments to explore the surface uniformly (Fig.\ \ref{fig_trapping}b). Unlike the low-density regime, where filaments form a single rotating band around the ``equator'' of the peanut-shaped surface, at higher \(Pe\) they move more uniformly without organizing into a coherent structure (Fig.~\ref{fig_trapping}d). They are no longer constrained to move collectively.
These results explain the distinct behavior of the Cassini oval with \(a=0.99b\) compared to those with \(a<0.99b\). Overall, these findings highlight how surface geometry can strongly influence filament dynamics and suggest that curvature and topology could be used to control or restrict the motion of active filaments or particles~\cite{paper-spin-connection}.

\section{Conclusions}
\label{sec:conclusions}
In this paper, we investigated the dynamics of both individual and collective active filaments confined to curved surfaces using numerical simulations. Our goal was to understand how surface curvature, filament bending energy, and activity interact to shape filament behavior.

For single filaments on axisymmetric surfaces such as Gaussian bumps, we found that filaments follow geodesic paths when activity dominates. In contrast, when bending rigidity is strong, filaments deviate from geodesics, as these no longer minimize their elastic energy. On spherical surfaces, where geodesics coincide with energy-minimizing configurations, filaments consistently follow great circles regardless of activity. When deforming the sphere into Cassini ovals, filaments continue to follow geodesic paths that also minimize bending energy, highlighting a robustness of geodesic guidance in this regime.

We introduced a geometry-dependent Péclet number, which captures how the timescale of filament motion depends on surface curvature. In dilute suspensions on Cassini ovals, individual filaments explore the full surface, while collectively they self-organize into rotating bands, typically aligned along the equator. For highly deformed geometries (e.g., a Cassini oval with \(a = 0.99b\)), we observed a transition: at low density, rotating bands still form, but at higher densities, filaments become trapped in one lobe and fail to explore the entire surface due to spatial crowding. A similar transition occurs as a function of activity: at low Péclet number, filaments localize, while higher activity enables full surface exploration. However, in this case, filaments exhibit more disordered motion, and coherent bands do not emerge.

Overall, our results demonstrate that surface curvature and topology can be leveraged to guide or constrain the behavior of active filaments \cite{paper-spin-connection}. This highlights the potential of geometric design as a mechanism to control active matter in soft materials, microfluidic devices, and bio-inspired systems \cite{Cereceda-LopezCurvature2024}.

\section{Acknowledgements}
E.D.M.\ acknowledges funding by the endowed E.N.\&M.N.\ Lindsay PhD studentship. R.S.\ acknowledges support from the UK Engineering and Physical Sciences Research Council (Award EP/W023946/1). D.M.F. and G.J. acknowledge support from the Comunidad de Madrid and the Complutense University of Madrid (Spain) through the Atracción de Talento program (Grant No. 2022-T1/TIC-24007). D.M.F. also acknowledges support from MINECO (Grant No. PID2023-148991NA-I00).
\begin{appendices}
\renewcommand{\thefigure}{A\arabic{figure}}
\renewcommand{\theequation}{A\arabic{equation}}
\setcounter{figure}{0}
\setcounter{equation}{0}

\section{Geodesic and normal curvature}
\label{geodesic_normal}
To understand the behavior of filaments moving on a Gaussian bump, we compute the normal and geodesic curvatures, as well as the geodesic torsion. When working with curves embedded in a surface, it is convenient to use the \emph{Darboux} frame. In this framework, we define the surface unit normal \(\mathbf{n}\), the tangent vector to the curve \(\mathbf{t}\), and the unit tangent normal $\mathbf{l} = \mathbf{t} \times \mathbf{n}$. 

The Darboux frame is characterized by structure equations analogous to the Frenet–Serret equations\cite{docarmo76}. These are,
\begin{equation}
\begin{split}
\frac{\mathrm{d}\mathbf{t}}{\mathrm{d}s} &= k_g\,\mathbf{l} + k_n\,\mathbf{n}, \\
\frac{\mathrm{d}\mathbf{l}}{\mathrm{d}s} &= -k_g\,\mathbf{t} + \tau_g\,\mathbf{n}, \\
\frac{\mathrm{d}\mathbf{n}}{\mathrm{d}s} &= -k_n\,\mathbf{t} - \tau_g\,\mathbf{l}.
\end{split}
\end{equation}
From these relations, the geodesic curvature, normal curvature, and geodesic torsion can be defined as
\begin{equation}
\begin{split}
k_g &= \mathbf{k} \cdot (\mathbf{t} \times \mathbf{n}),\\
k_n &= \mathbf{k} \cdot \mathbf{n},\\
\tau_g &= -\frac{\mathrm{d}\mathbf{n}}{\mathrm{d}s} \cdot (\mathbf{t} \times \mathbf{n}),
\end{split}
\end{equation}
where the curvature vector is defined by \(\mathbf{k} = \frac{\mathrm{d}\mathbf{t}}{\mathrm{d}s}\). 

In simulations, each filament is represented as a sequence of discrete points, and all spatial derivatives are approximated using finite‐difference schemes \cite{fornberg1988generation}. Specifically, the arc‐length increments are computed as
\begin{equation}
\Delta s_i = \left|\mathbf{r}_{i+1} - \mathbf{r}_i\right|,
\end{equation}
where \(\mathbf{r}_i\) denotes the position of the \(i^\textrm{th}\) point along the filament. To compute derivatives with respect to arc length \(s\), we use NumPy's\cite{harris2020array} \texttt{np.gradient} function, which applies second‐order accurate central differences.
\section{Gaussian Curvature: Cassini Ovals and Gaussian Bump}
\label{Gaussican_curv}
The Cassini surface is defined implicitly, \( f(x, y, z) = 0 \). For such implicitly defined surfaces, the Gaussian curvature \( K \) can be expressed in terms of the gradient \( \nabla f \) and the Hessian matrix \( H(f) \) as\cite{goldman2005curvature}
\begin{equation}
    K = -\frac{1}{|\nabla f|^4}\begin{vmatrix}
        H(f) & \nabla f^{T} \\
        \nabla f & 0
    \end{vmatrix}.
\end{equation}

We compute the curvature using \texttt{sympy}~\cite{10.7717/peerj-cs.103} to get
\begin{equation}
    K_a(x,y,z) = \frac{(a^2 + r^2)\left[a^2(2x^2 - r^2) + r^4\right]}{r^4\left[(a^2 + r^2)^2 - 4a^2x^2\right]},
\end{equation}
where \( r^2 = x^2 + y^2 + z^2 \).
In the special case \( a = 0 \), corresponding to a sphere of radius \( r \), this expression reduces to the expected result for a sphere, i.e., $K_{a=0}(x,y,z) = \frac{1}{r^2}$. 

For a Gaussian‑bump surface $z=h(r)$, the Gaussian curvature is
\begin{equation}
  K(r) = \frac{h''(r)\,h'(r)}{r\bigl(1 + h'(r)^2\bigr)^2}\,,
\end{equation}
where \(h'(r)\) and \(h''(r)\) denote the first and second derivatives of the height function \(h(r)\).
\section{Comparing Continuum and Discrete Bending Energy}
\label{discrete_vs_continuum}
\begin{figure}[t]
\includegraphics[width=\columnwidth]{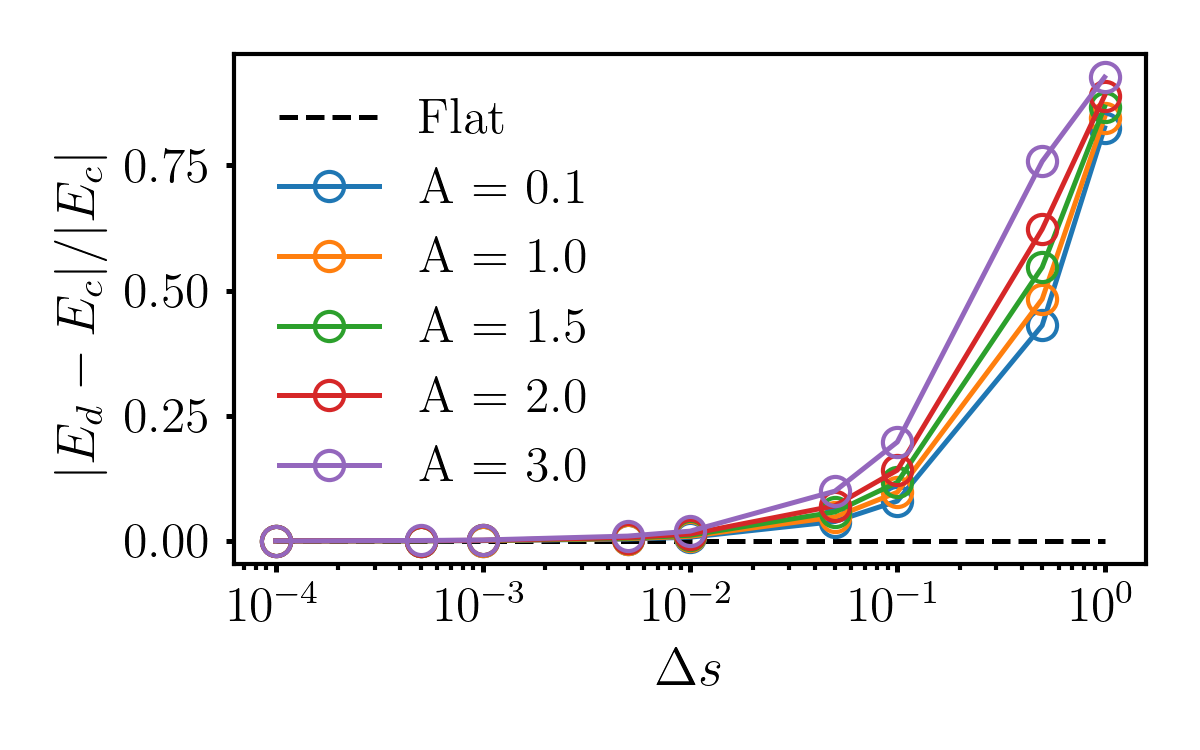}
\caption{\label{fig:error_energy} Error in the discrete bending energy \( E_d \) compared to the continuum bending energy \( E_c \), plotted as a function of monomer spacing \( \Delta s \), evaluated when the polymer head is located at the peak of the Gaussian bump (\( x = 0 \)). The dashed line is the case of a discrete polymer in a plane. while colored curves correspond to different bump amplitudes, \( A \).
}
\end{figure}

Figure~\ref{fig:error_energy} shows the relative error in the discrete bending energy $E_d$ compared to the continuum limit ($E_c$), for a filament positioned on top of the Gaussian bump. The continuum bending energy is computed by evaluating Eq.~\eqref{eq:cont-energy} numerically, using Simpson’s rule to approximate the integral\cite{press2007numerical}. As the monomer spacing $\Delta s$ increases, the error grows accordingly.

\section{Scaling Argument for Filament on Gaussian Bump}
\label{sec:scaling}
In Sec.~\ref{sec:single_active}, we saw through numerical results that depending on the value of $Pe$, when an active filament is propelled towards a Gaussian bump by its active propulsion, it may pass directly over, maintaining a geodesic trajectory, or deviate around the bump. In this section, we seek to understand this behavior by examining the competition between activity and bending rigidity through a simple scaling argument. 

We move to the overdamped limit of the filament dynamics and compare the time taken for the filament to pass over the bump, driven by its active self-propulsion $t_a$, with the time taken for the filament to deform due to its elastic bending forces $t_b$.

In the overdamped limit, where internal forces are typically balanced by external friction $\gamma v \approx f$. We can estimate the time to move over the entire filament over the bump when propelled by a constant force of magnitude $f_\mathrm{p}$ as $t_a \approx \frac{L \gamma}{f_\mathrm{p}}$. (Note this is valid as long as the size of the filament is much larger than that of the bump.) To compare this to $t_b$ we need to estimate the force caused by bending when passing over the bump. If we assume that the curvature angle induced by the bump is approximated by $\theta \approx \frac{\sigma_g}{A}$, then the bending energy penalty the filament pays for passing over the bump will be $E_b \approx \kappa \frac{\sigma_g^2}{A^2}$. The force on the filament is then given $f_b \approx \frac{E_b}{l_s}$, where $l_s$ is the length scale discussed in Sec.\ \ref{sec:simulation_model}, over which the filament senses the curvature. The filament is likely to experience this bending force until its path has deviated by a distance $l_s$, since this is approximately how far it must move to avoid sensing the bump. Therefore, the time taken minimize the bending penalty by avoiding the bump will be $t_b \approx \frac{l_s \gamma}{f_b} \approx \frac{\gamma l_s^2 A^2}{\kappa \sigma_g^2}=\frac{\gamma \sigma_g^2}{\kappa}$, since $l_s=\frac{\sigma_g^2} {A}$. 

This means whether the filament avoids or passes over the bump is quantified by the dimensionless number $\frac{t_b}{t_a} \approx \frac{f_\mathrm{p} l_s^2 A^2}{ L \kappa \sigma_g^2} \approx Pe \frac{l_s A^2}{L \sigma_g^2}$, where $Pe$ is the earlier defined geometry-dependent  Péclet number. Using the expression for $l_s$ for Gaussian bump, we have $\frac{t_b}{t_a} \approx Pe \frac{A}{L}$. This means for large $Pe$ we expect $t_a\ll t_b$, so the filament is expected to remain on a geodesic, consistent with our numerical results.
\begin{figure}[t]
\includegraphics[width=\columnwidth]{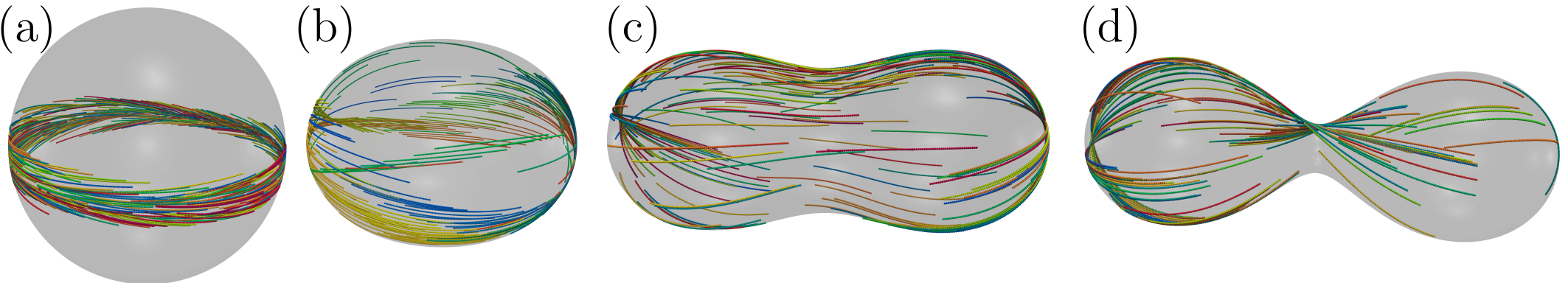}
\caption{Steady-state configurations of single filaments without noise at intermediate activity for four different values of the Cassini oval parameter aa: (a) Sphere ($a=0$); (b) prolate spheroid ($a=0.5b$); (c) Peanut-shaped surface ($a=0.9b$); and (d) Peanut-shaped surface with a narrow channel ($a=0.99b$). Each filament is initialized from the corresponding configuration shown in Fig.\ref{fig2}a-d.\label{fig:cassini_single_no_noise} 
}
\end{figure}

\begin{figure}[t]
\includegraphics[width=\columnwidth]{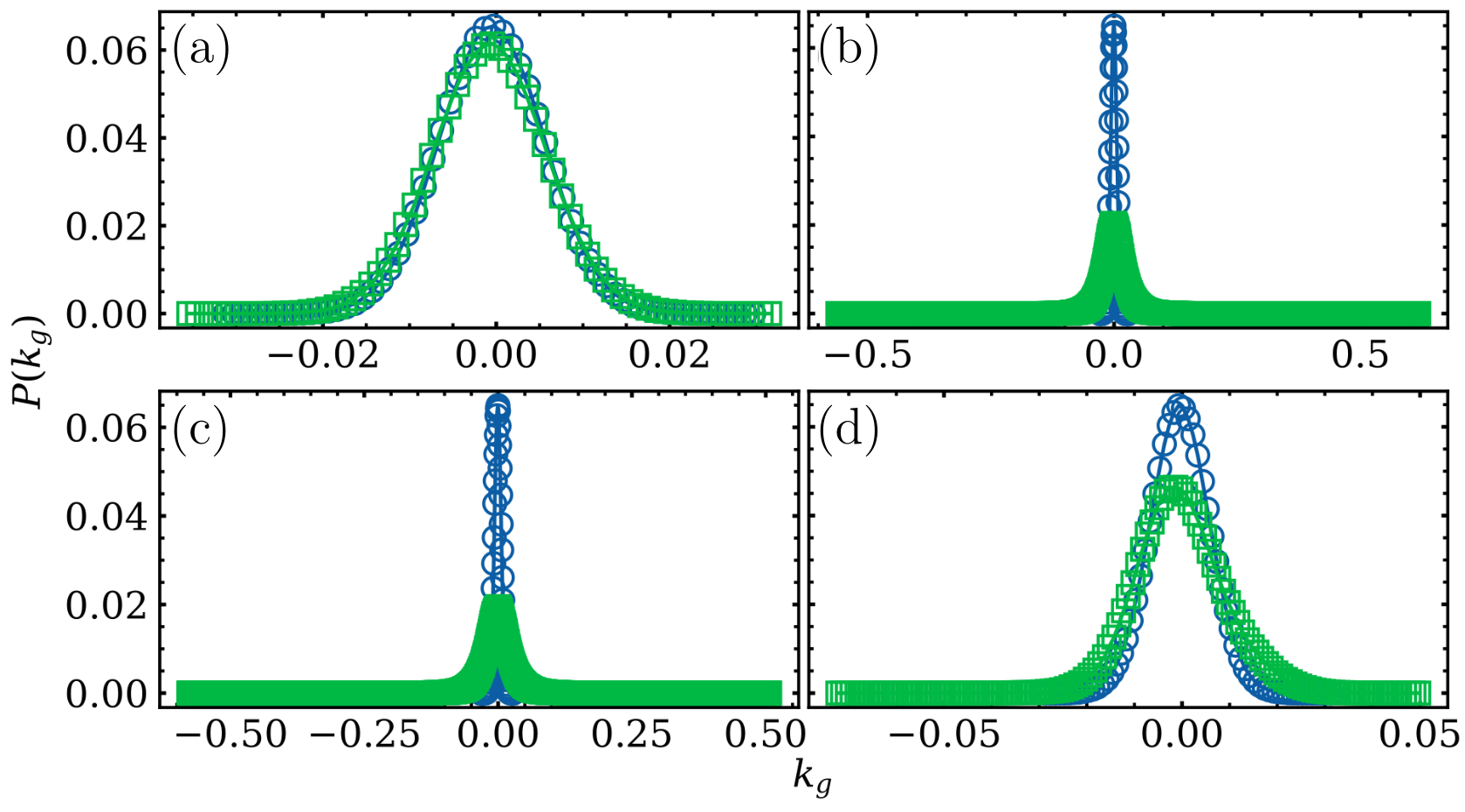}
\caption{Probability distributions of the geodesic curvature corresponding to the snapshots shown in Fig.~\ref{fig2}. Green squares represent the case of filament suspensions (Fig.\ \ref{fig2}a–d), while blue circles correspond to single filaments with noise (Fig.\ \ref{fig2}e–h). (a) sphere, (b) prolate spheroid, (c) peanut-shaped surface, and (d) peanut-shaped surface with a narrow channel.\label{fig:Probability distribution} 
}
\end{figure}
\section{Single Active filaments on Cassini ovals}
\label{Single_filament_on_Cassini}
Figure~\ref{fig:cassini_single_no_noise} shows the trajectories of single filaments without noise, initialized from the configurations shown in Fig.~\ref{fig2}a-d. In the spherical case (Fig.~\ref{fig:cassini_single_no_noise}a), filaments remain confined to the equator. In contrast, for nonzero values of the parameter $a$, the filaments progressively explore the entire surface and are no longer restricted to a specific region (Fig.~\ref{fig:cassini_single_no_noise}b–d).

Figure~\ref{fig:Probability distribution} presents the probability distributions of geodesic curvature corresponding to the configurations shown in Fig.~\ref{fig2}. For the spherical case (Fig.~\ref{fig2}a,e), the distributions for a single filament (Fig.~\ref{fig2}e) and for the filament suspension (Fig.~\ref{fig2}a) are nearly identical, both peaking around $k_g \approx 0$. Due to the presence of noise, the values of $k_g$ in the single-filament case are slightly higher than in the noiseless case. In contrast, for Cassini ovals with $a>0$, the probability of observing zero geodesic curvature is noticeably lower in the filament suspension than in the single-filament case (Fig.~\ref{fig2}b–d).

\end{appendices}
 \bibliographystyle{apsrev4-1} 
\bibliography{bib}
\end{document}